\documentclass[aps,reprint,pra,superscriptaddress,floats,floatfix]{revtex4-2}

\usepackage{amsmath}
\usepackage{amsfonts}
\usepackage{amssymb}
\usepackage{graphicx}
\usepackage{color}
\usepackage[colorlinks=true,citecolor=blue,linkcolor=blue,urlcolor=blue]{hyperref}
\usepackage{times}
\usepackage{amstext}
\usepackage{latexsym}
\usepackage[running,mathlines]{lineno}
\usepackage{verbatim}
\usepackage{physics}

\providecommand{\abs}[1]{\lvert#1\rvert}

\providecommand{\moy}[1]{\langle #1 \rangle}
\providecommand{\bra}[1]{\langle #1 \rvert}
\providecommand{\ket}[1]{\lvert #1 \rangle}

\newcommand{\I}{\mathrm{i}}

\begin{document}
\raggedbottom

\title{Superposition of two-mode squeezed states for quantum information processing and quantum sensing}

\author{Fernando R. Cardoso}
\email{frc@df.ufscar.br}
\affiliation{Departamento de F\'{i}sica, Universidade Federal de S\~{a}o Carlos, 13565-905 S\~{a}o Carlos, S\~{a}o Paulo, Brazil}

\author{Daniel Z. Rossatto}
\affiliation{Universidade Estadual Paulista (Unesp), Campus Experimental de Itapeva, 18409-010 Itapeva, S\~{a}o Paulo, Brazil}

\author{Gabriel P. L. M. Fernandes}
\affiliation{Departamento de F\'{i}sica, Universidade Federal de S\~{a}o Carlos, 13565-905 S\~{a}o Carlos, S\~{a}o Paulo, Brazil}

\author{Gerard Higgins}
\affiliation{Department of Physics, Stockholm University, SE-106 91 Stockholm, Sweden}

\author{Celso J. Villas-Boas}
\affiliation{Departamento de F\'{i}sica, Universidade Federal de S\~{a}o Carlos, 13565-905 S\~{a}o Carlos, S\~{a}o Paulo, Brazil}

\begin{abstract}
We investigate superpositions of two-mode squeezed states (TMSSs), which have potential applications to quantum information processing and quantum sensing. 
Firstly we study some properties of these nonclassical states such as the statistics of each mode and the degree of entanglement between the two modes, which can be higher than that of a TMSS with the same degree of squeezing. 
The states we consider can be prepared by inducing two-mode Jaynes-Cummings and anti-Jaynes-Cummings interactions in a system of two modes and a spin-$\tfrac{1}{2}$ particle, for instance in the trapped ion domain, as described here. We show that when two harmonic oscillators are prepared in a superposition of two TMSSs, each reduced single-mode state can be advantageously employed to sense arbitrary displacements of the mode in phase space. 
The Wigner function of this reduced state exhibits a symmetrical peak centered at the phase-space origin, which has the convenient peculiarity of getting narrower in both quadratures simultaneously as the average photon number increases. This narrow peak can be used as the pointer of our quantum sensor, with its position in phase space indicating the displacement undergone by the oscillator.
\end{abstract}

\maketitle

\section{\label{sec:1}Introduction}
In quantum mechanics, the superposition principle is the origin of fascinating nonclassical attributes of quantum states, such as quantum coherence \cite{Streltsov2017}, squeezing \cite{andrews2015photonics, *Lvovsky2014} and entanglement \cite{Horodecki2009}. Great efforts, both theoretical and experimental, have been made in order to generate nonclassical states, and to investigate their properties.
For instance, entangled states can be used to evade the standard formulation of Heisenberg's uncertainty principle \cite{Colangelo2017}.
These states have a wide range of applications in quantum information processing \cite{Nielsen, Ladd2010, Terhal2015, Gisin2007, Scarani2009, Kimble2008, Meter2014}, quantum-enhanced metrology \cite{Giovannetti2004,Giovannetti2011,Tth2014,Pezze2018} and fundamental tests of quantum mechanics \cite{Brunner2014,Rossatto2011}. 

Consider the two-mode squeezed state (TMSS), which is an entangled state of two bosonic modes \cite{andrews2015photonics, *Lvovsky2014}. This kind of correlation exhibits Bell nonlocality \cite{Brunner2014}, a key ingredient to demonstrate the Einstein-Podolsky-Rosen (EPR) paradox \cite{einstein1935,Reid2009} (fundamental test of quantum mechanics), to implement quantum teleportation in continuous variables \cite{Pirandola2015} (manipulation of quantum information) and to detect very weak fields \cite{Anisimov2010} such as the gravitational waves \cite{Schnabel2017} (quantum metrology). Another intriguing nonclassical state is the so-called cat state, a quantum superposition of two diametrically opposed coherent states, which has been employed to demonstrate Schrödinger's famous cat paradox and has been used as a resource for quantum information processing \cite{Schleich1991,Sanders1992,Enk2001,Jeong2001,Wenger2003,Jeong2003,Ralph2003,Gilchrist2004,Magdalena2007,Lund2008,Leghtas2013,Wineland2013,Haroche2013,Vlastakis2013,Mirrahimi2014,Wang2016,Ofek2016,Albert2018,Hacker2019}.

Here we are interested in nonclassical states that connect the concepts of both TMSSs and cat states, namely a superposition of TMSSs with same       amplitudes but opposite phases. There are already theoretical proposals for their generation in microwave cavities \cite{villas-boas2005} (that could also be adapted for the context of optical cavities or solid-state-based systems), trapped ions \cite{Zeng2002} and by using the parity operator \cite{karimi2018}. Nevertheless, as far as we know, the literature lacks a study on their properties and applications, which is our goal here.

We first analyze the statistics of each reduced single-mode state (RSMS) of a superposition of TMSSs. For the symmetrical balanced superposition, which we call the even TMSS, we show that there is bosonic superbunching \cite{Ficek,Paul1982}, an effect with potential applications for advanced imaging techniques (such as ghost interference and imaging) as well as efficient nonlinear light-matter interaction \cite{Bhatti2015,Bai2017,Zhou2017,Lettau2018,Liu2018,Marconi2018,Zhang2019}. On the other hand, for small squeezing parameters, each single mode of the asymmetrical balanced superposition (odd TMSS) presents two-photon anticorrelation \cite{Ficek,Paul1982}, a desired behavior for single-photon sources \cite{Lounis2005}. Remarkably, each mode of the even and odd TMSSs behave as pseudo-thermal states, which consist of thermal states with only even and odd Fock excitations, respectively. In addition, we investigate the entanglement degree between the modes of such cat-like states, which can be higher than that of the TMSS in certain parameter regimes. Since entanglement is a resource of quantum states, this may be an advantage for quantum information processing.



Afterwards, we show that the RSMS of either even or odd TMSS can be used to sense the amplitude of arbitrary single-mode displacements acting on a harmonic oscillator. The Wigner functions of these RSMSs each have a symmetrical peak centered at the phase-space origin, which gets narrower in both quadratures simultaneously as the squeezing parameter and the average number of excitations in the state increases, but without violating Heisenberg's uncertainty relation. This narrow peak works as the pointer of our quantum sensor, with its position in the phase space indicating the displacement undergone by the oscillator, which could physically describe, for instance, an optical \cite{Hacker2019} or a microwave resonator mode \cite{Vlastakis2013,Penasa2016,Federov2010}, vibrational modes of trapped ions \cite{Zeng2002,Blatt2003} or a nanomechanical oscillator \cite{Regal2008}. In this sense, our sensor is able to probe any time-dependent classical force inducing a displacement on a quantum resonator \cite{Terhal2017}.

Several studies attempt to figure out the ultimate limits of measuring forces and displacements on an oscillator \cite{Caves1980}, beating the standard quantum limit and even reaching the Heisenberg one \cite{Giovannetti2004,Penasa2016,Facon2016,Munro2002,Didier2015,Schreppler2014,Mason2019,Wang2019,Zwierz2010}. However, it worth stressing that this is not our goal, especially because the saturation of the conventional bound for the Heisenberg limit, which is derived directly from the properties of the quantum Fisher information, has been recently called into question \cite{Wiseman2020}. Hence, as similarly considered in Ref.~\cite{Terhal2017}, which described the determination of both parameters of a displacement acting on an oscillator, the idea we put forward here is the possibility of sensing displacements undergone by a quantum resonator with a single-mode sensor state robust against phase errors when measuring a phase-space quadrature.
Furthermore, our results hold regardless the displacement strength, not being limited to small amplitudes as is the case for grid states \cite{Terhal2017}.

The paper is organized as follows: Section \ref{sec:2} outlines the procedure for generating superpositions of TMSSs by coupling a two-level quantum system with two bosonic field modes, and presents the expressions for the two mode superposition and the reduced density matrices. In Sec.~\ref{sec:3}, we present the relevant statistical properties for those states, such as Wigner functions and populations in the Fock basis, and also the second-order correlation function $g^{(2)}(0)$, remarking properties of anti and superbunching, and show that odd TMSS can be used as a source of single photons in two modes. Section~\ref{sec:4} provides the entanglement properties of the superpositions of TMSSs and shows that, for certain regime of parameters, the superpositions of TMSSs show more entanglement than TMSSs. In Sec.~\ref{sec:5} we present basic concepts of quantum metrology and discuss potential applications of even and odd TMSSs in detecting small coherent forces in any direction, (they are sensitive to displacements in all directions in phase space). In Sec~\ref{sec:6}, a simulation of the process for generating even and odd TMSSs in the trapped-ion domain is presented. The process involves coupling the electronic state of the ion with two of its motional degrees of freedom using a two-colour laser field. We also discuss the process of probing the Wigner function. Finally, Section~\ref{sec:7} summarises the results and presents the conclusions.

\section{\label{sec:2}Generation of superposition of two-mode squeezed states}
TMSSs can be generated by coupling two bosonic modes with a two-level quantum system (a qubit), via the Hamiltonian ($\hbar = 1$)
\begin{align}
\mathcal{H}&=-(\chi^{\ast}ab+\chi a^{\dagger}b^{\dagger})\sigma_{x} \nonumber \\
&= -(\chi^\ast a b \sigma^+ + \chi a^\dagger b^\dagger \sigma^-) -(\chi^{\ast} a b \sigma^- + \chi a^\dagger b^\dagger \sigma^+) \label{eq:heff4},
\end{align}
where $\chi$ is the coupling strength, $a$ ($a^{\dagger}$) and $b$ ($b^{\dagger}$) are the annihilation (creation) operators for the bosonic modes, and $\sigma_{x}=\ket{+}\bra{+}-\ket{-}\bra{-}$ is the Pauli-X operator, with $\ket{\pm}=\frac{1}{\sqrt{2}}\left(\ket{\rm g}\pm\ket{\rm e}\right)$, where $\ket{\rm e}$ ($\ket{\rm g}$) is the excited (ground) state of the two-level system, and $\sigma^\pm=\frac{1}{2}(\sigma_x\pm i \sigma_y)$ are the fermionic raising and lowering operators. 
The two terms in the second line of Eq.~\ref{eq:heff4} are a two-mode Jaynes-Cummings interaction and a two-mode anti-Jaynes-Cummings interaction.

The coupling Hamiltonian $\mathcal{H}$ can be realized in various platforms. In Sec.~\ref{sec:6} we describe how it may be implemented in a trapped-ion setup.

We consider the case where the two-level system is initially in the superposition state $\ket{\phi_{0}}=\frac{1}{\sqrt{2}}\left(\ket{-}+e^{\I\varphi}\ket{+}\right) \equiv (\cos(\varphi/2)\ket{\rm g} + \I\sin{(\varphi/2)}\ket{\rm e})$ while the two bosonic modes are each in the vacuum state $\ket{\psi_{0}}=\ket{0,0}$. Thus the initial state of the composite system is separable $\ket{\Psi_0}=\ket{\phi_{0}}\ket{\psi_{0}}$ (the tensor product symbol is omitted for brevity).
After applying the coupling $\mathcal{H}$ for time $\tau$ the composite system evolves to
\begin{equation}
\ket{\Psi_\tau} ={e}^{-\I\mathcal{H}\tau}\ket{\Psi_0} =\tfrac{1}{\sqrt{2}}(\ket{-}\ket{\psi(\xi)}+{e}^{\I\varphi}\ket{+}\ket{\psi(-\xi)}),\label{eq: psi_tau}
\end{equation}
where
\begin{align}
\begin{split}
\ket{\psi(\xi)} &=\mathrm{{e}}^{\left(\xi^{\ast} ab-\xi a^{\dagger}b^{\dagger}\right)}\ket{0,0} \\
&={ \frac{1}{\cosh(r)}\sum_{n=0}^{\infty}\left[- e^{\I \theta}\tanh(r)\right]^{n}\ket{n,n}}
\end{split}
\end{align}
is the TMSS which we parameterize by $\xi = -\I\chi\tau=re^{\I \theta}$, with squeezing parameter $r = |\chi|\tau$ and squeezing angle $\theta = \arg{(\xi)}$ \cite{gerryknight}. 

From Eq.~\eqref{eq: psi_tau} the bosonic modes are projected onto a TMSS by measurement of the two-level system in the X-basis $\{\ket{+},\ket{-}\}$.
Starting from the TMSS, the reduced density matrix of one mode (found by tracing out the variables of the other mode) is a thermal state
\begin{equation}
    \rho_\text{th}=(1-\lambda_r)\sum_{n=0}^{\infty}\lambda_{r}^{n}\dyad{n}=\sum_{n=0}^{\infty}\tfrac{\moy{n}_\text{th}^n}{(1+\moy{n}_\text{th})^{n+1}}\dyad{n}, \label{rho_th}
\end{equation}
with average number of excitations $\moy{n}_\text{th}=\Tr(a^\dagger a \rho_\text{th})=\sinh^{2}(r)=\tfrac{\lambda_r}{1-\lambda_r}$ and $\lambda_r = \tanh^{2}(r)$ \cite{gerryknight}.

More interesting results emerge when the two-level system is projected onto the Z-basis $\{\ket{\rm e},\ket{\rm g}\}$. In this basis Eq.~\eqref{eq: psi_tau} becomes
\begin{equation} \label{eq: psi_general0}
    \ket{\Psi(\xi,\varphi)}=\tfrac{1}{2\mathcal{N}_{+}}   \ket{\rm g} \ket{\psi_{+}(\xi,\varphi)} - \tfrac{1}{2\mathcal{N}_{-}} \ket{\rm e} \ket{\psi_{-}(\xi,\varphi)},
\end{equation}
where
\begin{equation} \label{eq: psi_general}
\ket{\psi_{\pm}(\xi,\varphi)}= \mathcal{N}_{\pm}(\ket{\psi(\xi)} \pm {e}^{\I\varphi}\ket{\psi(-\xi)})    
\end{equation}
are superposition states of two diametrically opposed TMSSs, with $\vert \mathcal{N}_{\pm} \vert^{2} = {\tfrac{1}{2}\tfrac{1+\lambda_r}{(1+\lambda_r)\pm\epsilon_\varphi (1-\lambda_r)}}$ and $\epsilon_\varphi = \cos{\varphi}$. When the two-level system is projected onto $\ket{\rm g}$ ($\ket{\rm e}$) the bosonic modes are projected onto the cat-like state $\ket{\psi_{+}(\xi,\varphi)}$ ($\ket{\psi_{-}(\xi,\varphi)}$). Since $\varphi \in [0,2\pi)$ and $\ket{\psi_{+}(\xi,\varphi)} = \ket{\psi_{-}(\xi,\varphi+\pi)}$, it is sufficient to analyze just the properties of one of these states, e.g., $\ket{\psi_{+}(\xi,\varphi)}$. We refer to the states that emerge when $\varphi=0$ and $\varphi=\pi$ as even and odd TMSSs respectively, $\ket{\psi_{E}(\xi)} \equiv \ket{\psi_{+}(\xi,0)} = \ket{\psi_{-}(\xi,\pi)}$ and $\ket{\psi_{O}(\xi)} \equiv \ket{\psi_{+}(\xi,\pi)} = \ket{\psi_{-}(\xi,0)}$, because they comprise only even and odd bosonic excitation
\begin{align}
\ket{\psi_{E}(\xi)}&=\sqrt{{1-\lambda_{r}^{2}}}\sum_{n=0}^{\infty}(-\lambda_{r}^{\frac{1}{2}}e^{\I\theta})^{2n}\!\ket{2n,2n}, \label{eq: psi_even} \\
\ket{\psi_{O}(\xi)}&=\sqrt{\frac{1\!-\!\lambda_{r}^{2}}{\lambda_{r}}}\sum_{n=0}^{\infty}(-\lambda_{r}^{\frac{1}{2}}e^{\I\theta})^{2n+1}\!\ket{2n\!+\!1,2n\!+\!1}. \label{eq: psi_odd}
\end{align}

When the two modes are in the state $\ket{\psi_{+}(\xi,\varphi)}$ the reduced density matrix of each mode is
\begin{equation}
\rho(r,\varphi)=2(1\!-\!\lambda_r)\abs{\mathcal{N}_{+}}^{2}\sum_{n=0}^{\infty}\lambda_{r}^{n}\left[1\!+\!(-1)^{n} \cos{\varphi} \right]\dyad{n} \label{eq: rho_single_general},
\end{equation}
which is independent of the squeezing angle $\theta$. Specifically for the even and odd TMSSs we have the RSMSs
\begin{align}
\rho_{E}&=(1-\lambda_{r}^{2})\sum_{n=0}^{\infty}\lambda_{r}^{2n}\dyad{2n}, \label{eq: rho_even} \\
\rho_{O}&=(1-\lambda_{r}^{2})\sum_{n=0}^{\infty}\lambda_{r}^{2n}\dyad{2n+1}. \label{eq: rho_odd}
\end{align}
As the even and odd TMSSs [Eqs.~\eqref{eq: psi_even}-\eqref{eq: psi_odd}] are built from the superposition of TMSSs, whose reduced single-mode states are described by thermal states [Eq.~\eqref{rho_th}], it is not surprising that the reduced single mode of the even and odd TMSSs behave as `pseudo-thermal states'. We can indeed identify that by rewriting Eqs.~\eqref{eq: rho_even}-\eqref{eq: rho_odd} in terms of $\moy{n}_\text{th}$ and comparing them with $\rho_\text{th}$, namely
\begin{align}
\rho_{E}&=\frac{1\!+\!2\moy{n}_\text{th}}{1\!+\!\moy{n}_\text{th}}\sum_{n=0}^{\infty}\dfrac{\moy{n}_\text{th}^{2n}}{(1+\moy{n}_\text{th})^{2n\!+\!1}}\dyad{2n}, \label{eq: rho_even2} \\
\rho_{O}&=\frac{1\!+\!2\moy{n}_\text{th}}{\moy{n}_\text{th}}\sum_{n=0}^{\infty}\dfrac{\moy{n}_\text{th}^{2n\!+\!1}}{(1\!+\!\moy{n}_\text{th})^{(2n\!+\!1)\!+\!1}}\dyad{2n\!+\!1}. \label{eq: rho_odd2}
\end{align}

From the above expressions one can recognize $\rho_{E}$ and $\rho_{O}$ as even and odd thermal states (`pseudo-thermal states') respectively ~\cite{dodonov2003theory}, which are particular cases of binomial negative states \cite{Obada1997}. Even thermal state (even RSMS) can also be generated through a parametric pumping field with fluctuations \cite{Abdala2001}.

\section{\label{sec:3}Statistical properties}

\begin{figure}[t]
\includegraphics[trim = 2mm 1mm 2mm 2mm,clip,width=0.495\linewidth]{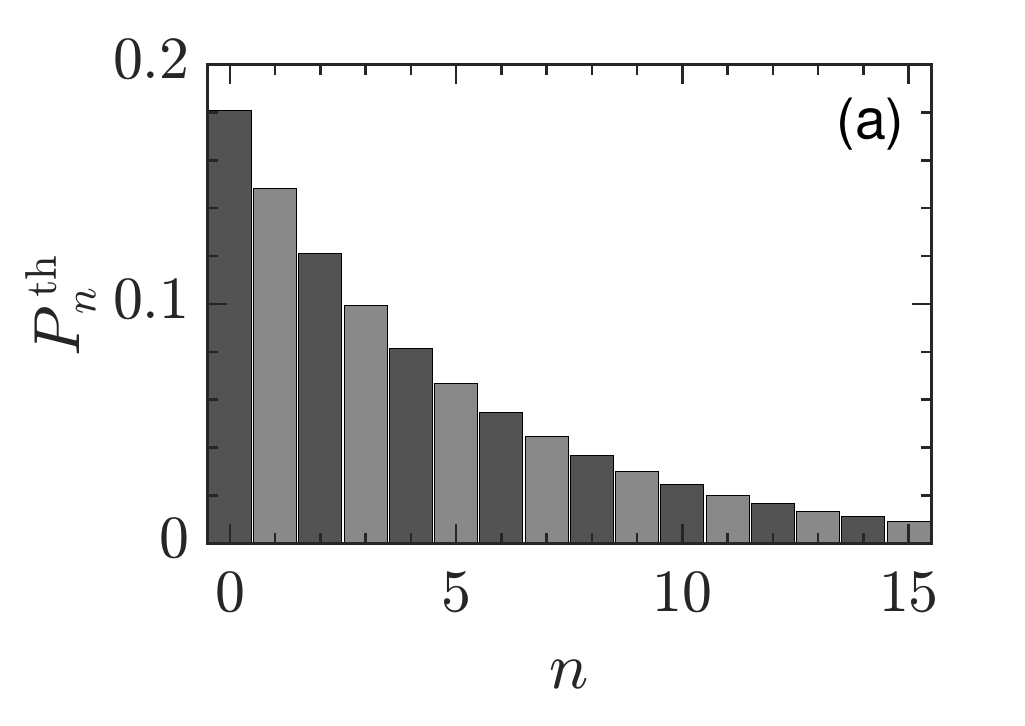} 
\includegraphics[trim = 2mm 0mm 7.75mm 12mm,clip,width=0.495\linewidth]{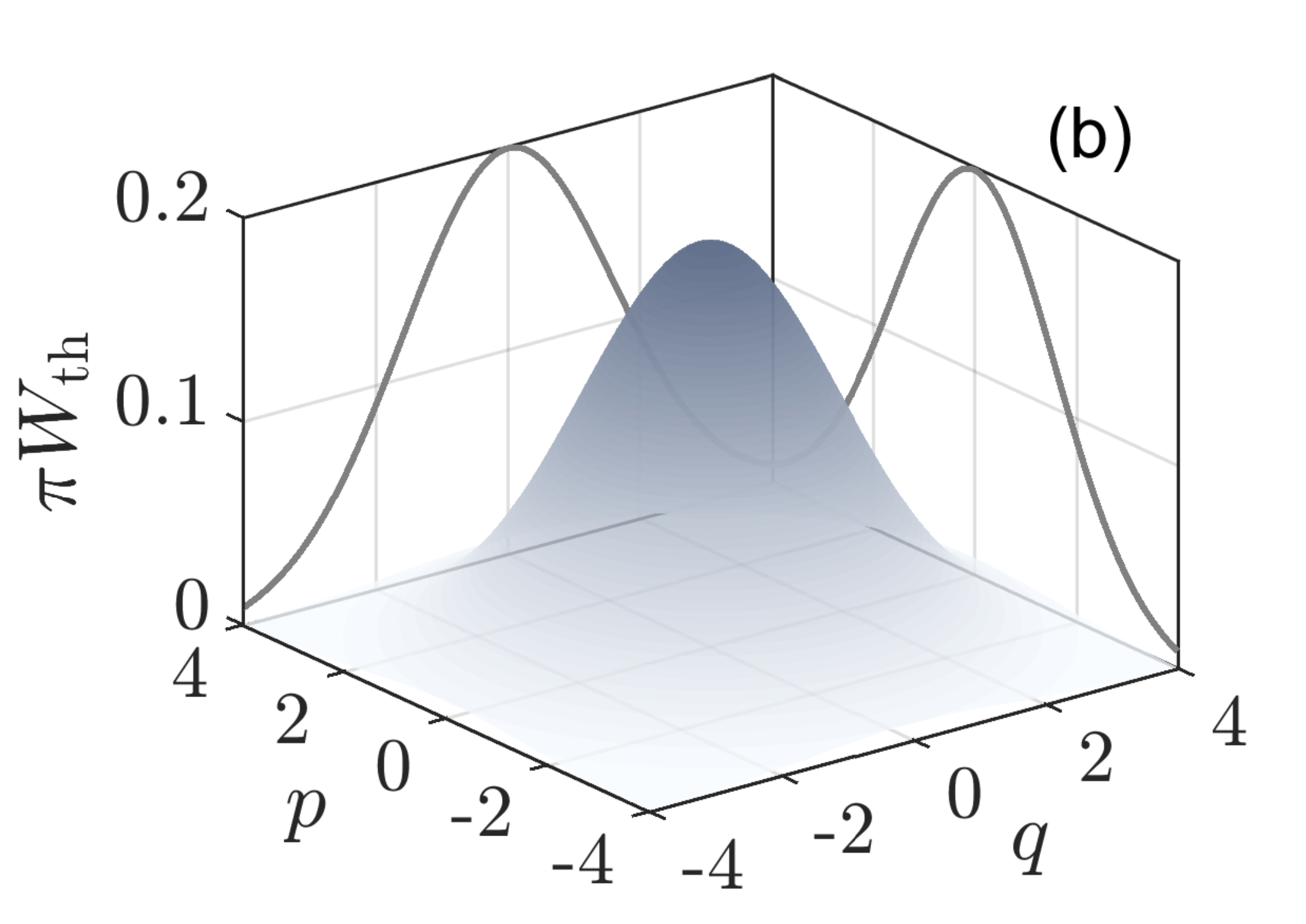} \\
\includegraphics[trim = 2mm 1mm 2mm 2mm,clip,width=0.495\linewidth]{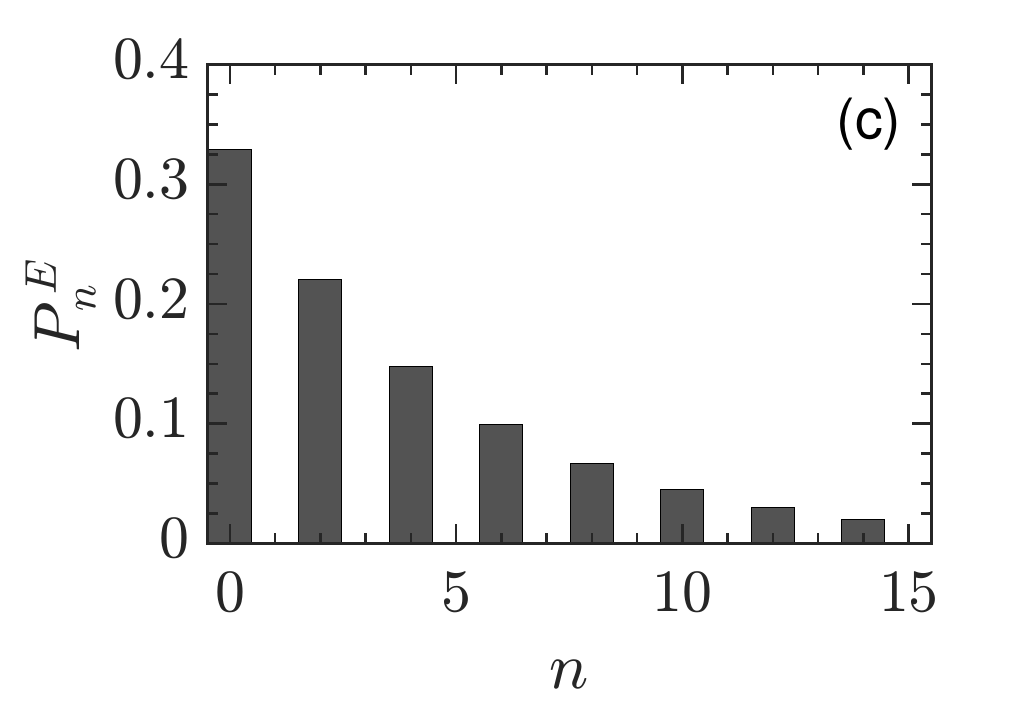}
\includegraphics[trim = 2mm 0mm 7.75mm 12mm,clip,width=0.495\linewidth]{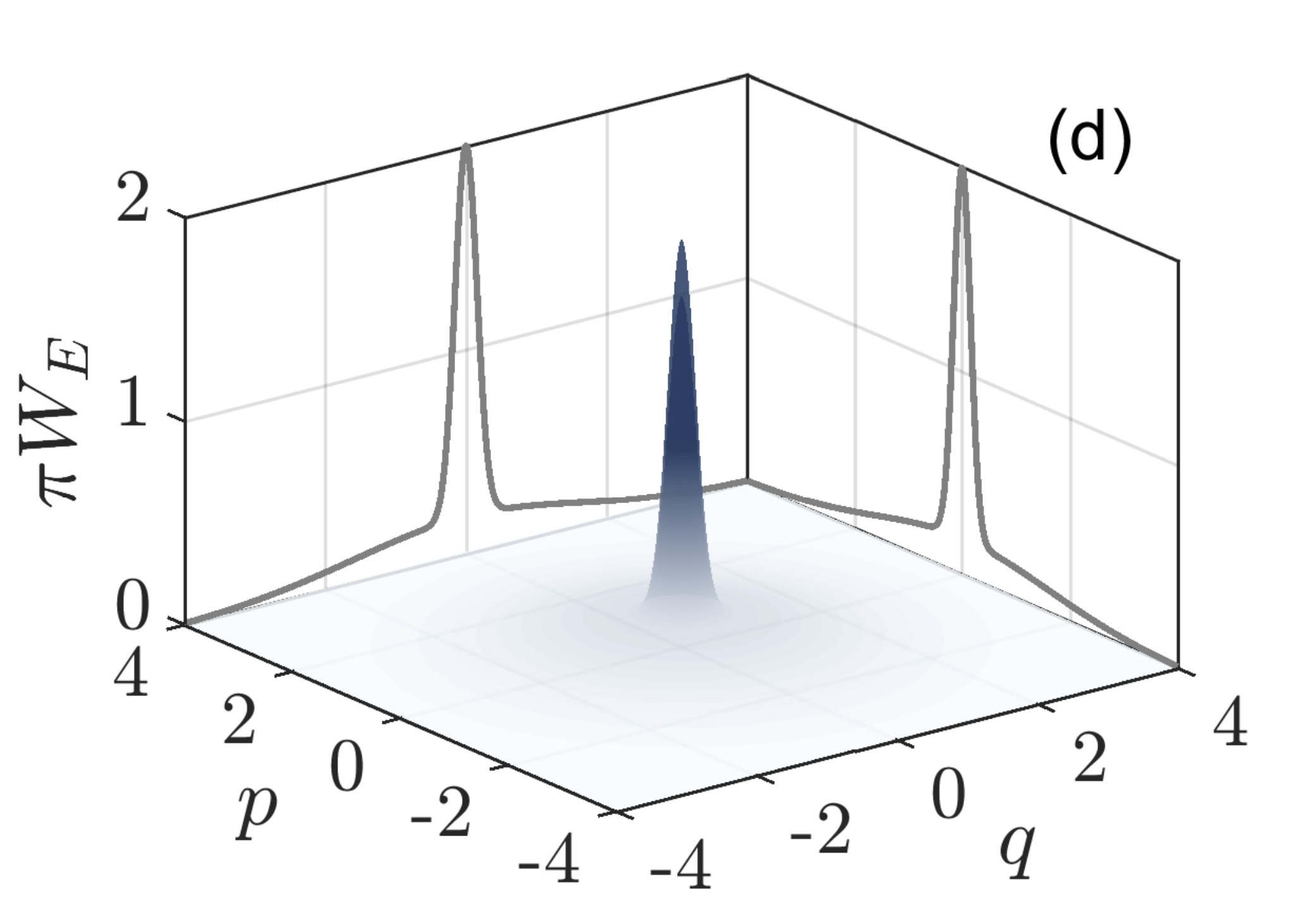} \\
\includegraphics[trim = 2mm 1mm 2mm 2mm,clip,width=0.495\linewidth]{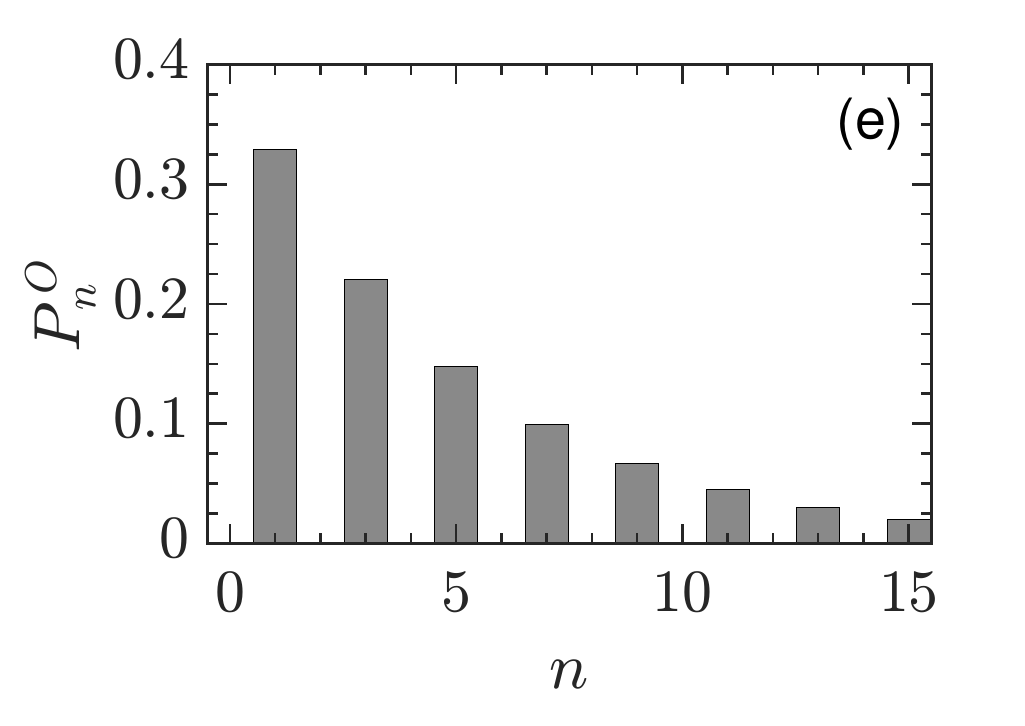}
\includegraphics[trim = 2mm 0mm 7.75mm 12mm,clip,width=0.495\linewidth]{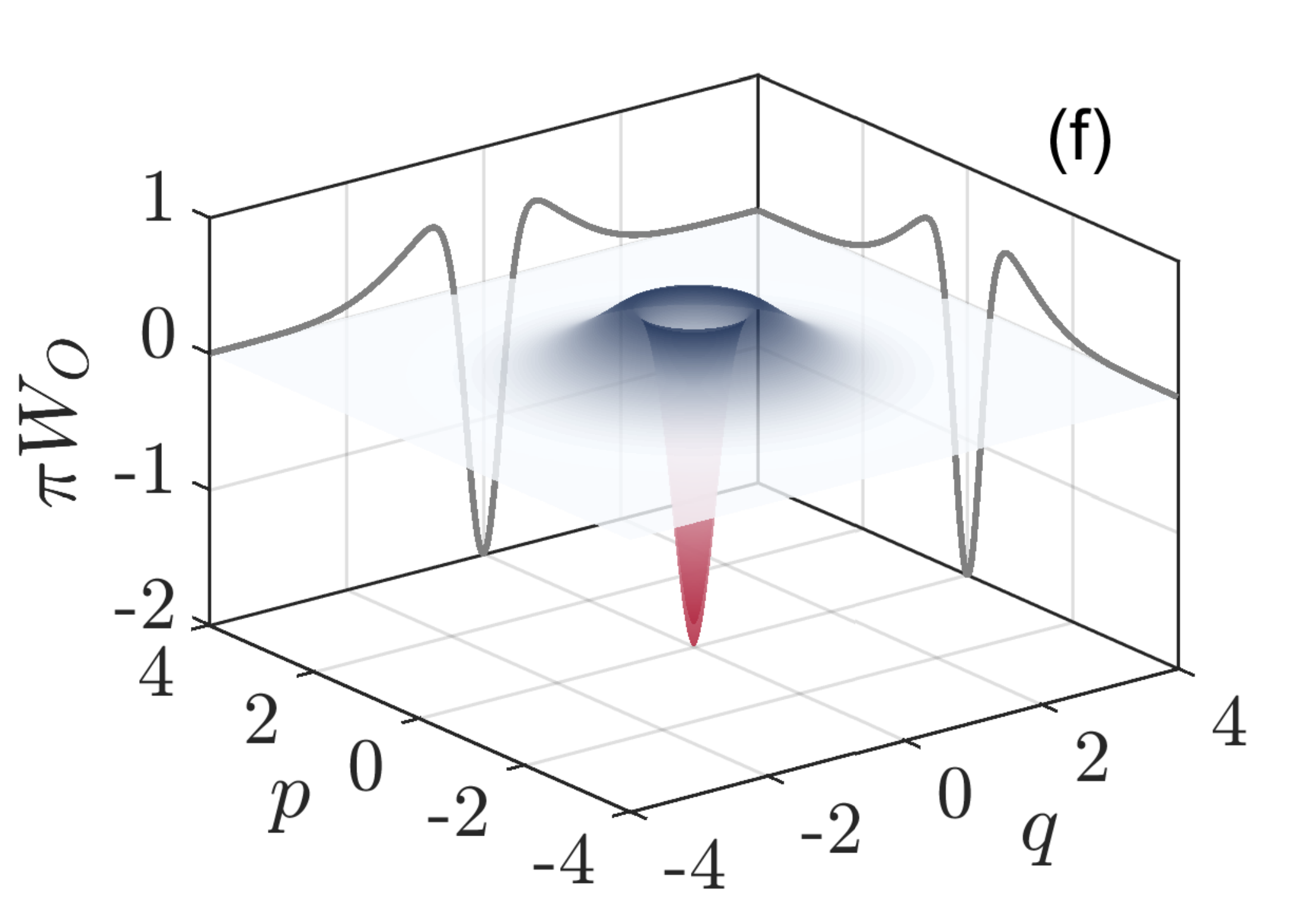}
\caption{\label{Pn_Wigner} Statistical properties of the reduced single-mode states. In the left panels we show the populations in the Fock basis of the (a) thermal state [$P_{n}^\text{th}$] and for each mode of the (c) even [$P_{n}^{E}$] and (e) odd [$P_{n}^{O}$] TMSSs for squeezing parameter $r=1.5$. In the right panels (b), (d) and (f) we show the Wigner functions for the same reduced single-mode states, respectively.}
\end{figure}

Let us discuss the properties of each single mode of the cat-like states. Since $\rho(r,\varphi)$ is diagonal in the Fock basis, with populations  $P_{n}( \varphi)=2(1-\lambda_r)\abs{\mathcal{N}_{+}}^{2}\lambda_{r}^{n}\left[1+(-1)^{n} \cos{\varphi} \right]$, the Wigner function of this state can be written as $W(q,p)=\sum_{n=0}^{\infty}P_{n}W_{n}(q,p)$, in which $W_{n}(q,p)=(2/\pi)\left(-1\right)^{n}L_{n}\!\left(4s^{2}\right)e^{-2s^{2}}$ is the Wigner function of the Fock state $\ket{n}$, 
with the Laguerre
polynomial $L_{n}\!\left(x\right)$ and $s^{2}=q^{2}+p^{2}$. Here $q$ and $p$ are, respectively, the eigenvalues of the position and momentum quadrature operators of the mode, $\hat{q}=a+a^{\dagger}$ and $\hat{p}=\I(a^{\dagger}-a)$, representing the dimensionless amplitudes of the mode quadratures in phase space \cite{walls2007quantum}. 

First we observe that $\rho(r,\varphi)$ reduces to the thermal state $\rho_\text{th}$ when $\varphi=\{\tfrac{\pi}{2},\tfrac{3\pi}{2}\}$ because $\ket{\psi_{+}(\xi,\varphi = \tfrac{\pi}{2}, \tfrac{3\pi}{2})}$ reduces to $\ket{\psi(\xi)}$ except for a global phase factor. The Wigner function of the thermal state $\rho_\text{th}$ is given by the two-dimensional Gaussian function
\begin{equation}
    W_\text{th}(q, p) = \dfrac{2}{\pi}\dfrac{e^{-2\frac{(q^2+p^2)}{2\moy{n}_\text{th} +1}}}{2\moy{n}_\text{th} +1} = \dfrac{2}{\pi}\dfrac{1-\lambda_{r}}{1+\lambda_{r}}e^{-2\frac{1-\lambda_{r}}{1+\lambda_{r}}(q^2+p^2)}, \label{Wthermal}
\end{equation}
while for each single mode of the even ($\rho_{E}$) and odd ($\rho_{O}$) TMSSs the Wigner functions are each described by sums of two two-dimensional Gaussian functions
\begin{align}
W_{\substack{E\\O}}(q, p)= \varpi_{\substack{E\\O}} &\left[  (1-\lambda_{r})e^{-2\left(\tfrac{1-\lambda_{r}}{1+\lambda_{r}}\right)(q^2+p^2)} \right. \nonumber \\ 
 &\left. \pm (1+\lambda_{r})e^{-2\left(\tfrac{1+\lambda_{r}}{1-\lambda_{r}}\right)(q^2+p^2)} \right], \label{eq: W_even_odd}
\end{align}
with $\varpi_{E} = \pi^{-1}$ and $\varpi_{O} = (\pi \lambda_{r})^{-1}$. 

Figure~\ref{Pn_Wigner} shows the populations and the Wigner function for $\rho_\text{th}$ [$P_{n}^\text{th}$ and $W_\text{th}$], $\rho_{E}$ [$P_{n}^{E}$ and $W_{E}$] and $\rho_{O}$ [$P_{n}^{O}$ and $W_{O}$] for squeezing parameter $r=1.5$. We observe that the profile of the populations for $\rho_{E}$ and $\rho_{O}$ are very similar, namely $P_{2n+1}^{O} = P_{2n}^{E} = (1-\lambda_{r}^2)\lambda_{r}^{2n}$, that is the probability distribution of $\rho_{O}$ is shifted by one unit compared with the distribution of $\rho_{E}$.

Accordingly the average number of excitations are related by $\moy{n}_{O} = \moy{n}_{E} + 1$, where $\moy{n}_{E}=\Tr(a^\dagger a \rho_E) = 2\lambda_{r}^2/(1-\lambda_{r}^2)$. Moreover, we notice that $P_{n}^{E}$ and $P_{n}^{O}$ are very similar to $P_{n}^\text{th}$, except for a normalization factor and for being nonzero only for even and odd Fock numbers, respectively; which illustrates the `pseudo-thermal' behavior of each mode of the even and odd TMSSs. It is also important to notice from Fig.~\ref{Pn_Wigner} a concentrated profile of the Wigner functions around the phase space origin for both $\rho_{E}$ and $\rho_{O}$, which can be useful for metrology purposes, as we discuss in Sec.~\ref{sec:5}.

\begin{figure}[t]
\includegraphics[trim = 2mm 2.5mm 2mm 0mm,clip,width=0.495\linewidth]{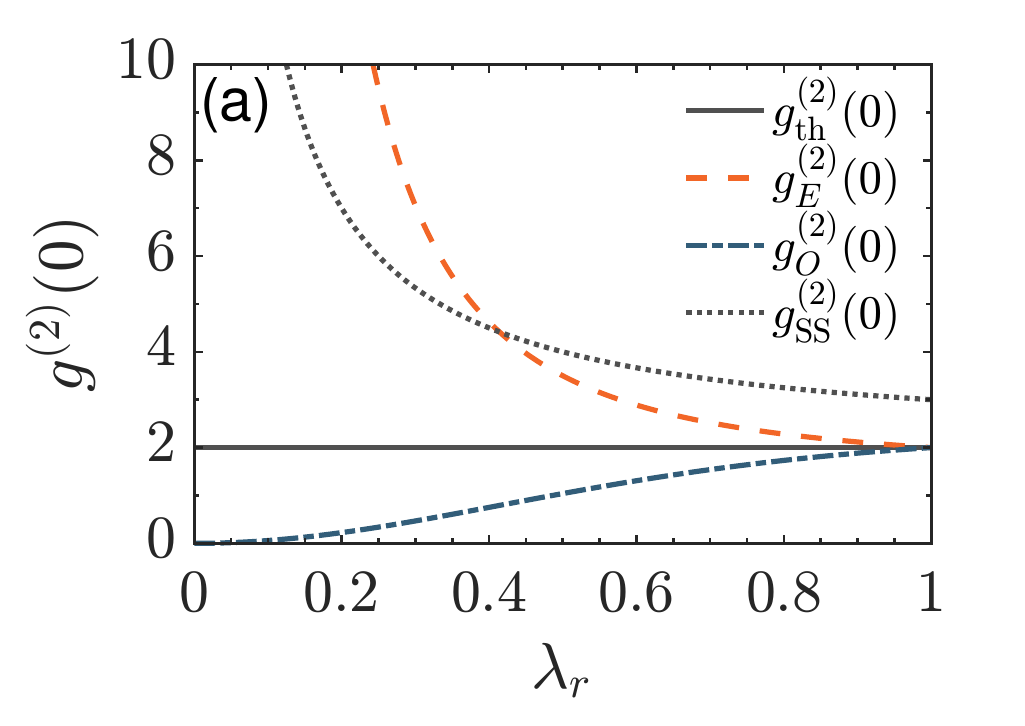}
\includegraphics[trim = 2mm 2.0mm 2mm 0mm,clip,width=0.495\linewidth]{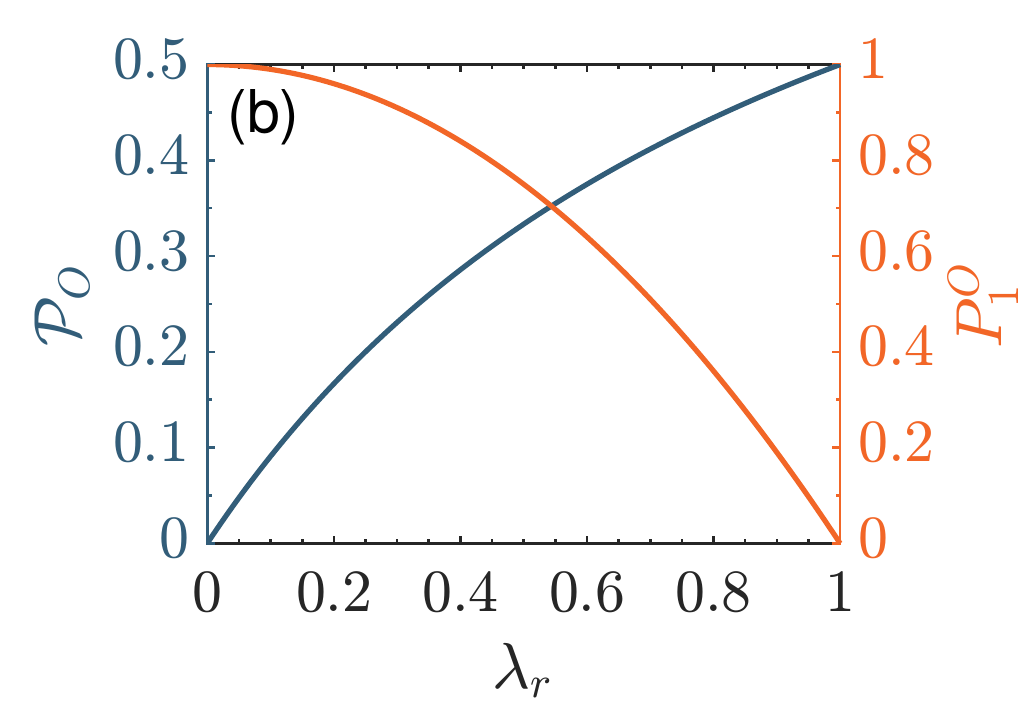}
\caption{\label{fig:Second-order-correlation}(a)~Second-order correlation function at zero-time delay
$g^{(2)}(0)$ as a function of the squeezing parameter $r$, via $\lambda_r = \tanh^{2}(r)$, for the thermal state ($\rho_\text{th}$), the RSMSs of the even ($\rho_{E}$) and odd ($\rho_{O}$) TMSSs, and the single-mode squeezed state ($\ket{\rm SS}$). (b)~Probability of projecting onto the odd TMSS state ($\mathcal{P}_{O}$) and probability of it having a single-photon in each mode ($P_{1}^{O}$) as a function of $\lambda_r$.
}
\end{figure}

We can also investigate the statistics of $\rho_{E}$ and $\rho_{O}$ using the second-order correlation function at zero-time delay $g^{(2)}(0)=\moy{a^{\dagger}a^{\dagger}aa}/\moy{a^{\dagger}a}^{2}$,
\begin{align}
\rho_{E}: \quad g_{E}^{(2)}(0)&= 2 + \dfrac{1-\lambda_{r}^2}{2\lambda_{r}^2} \ge  2, \label{eq: g20p} \\
\rho_{O}: \quad g_{O}^{(2)}(0)&= 2 - \dfrac{2(1-\lambda_{r}^2)}{(1+\lambda_{r}^2)^2} \le  2. \label{eq: g20m}
\end{align}
For large values of squeezing parameter ($\lambda_r \to 1$) we observe that the statistics of $\rho_{E}$ and $\rho_{O}$ tends to the thermal one [$g_\text{th}^{(2)}(0)=2$], i.e., these states become statistically indistinguishable when $r\gg 1$.
On the other hand, when the squeezing parameter is small, the RSMSs of the even and odd TMSSs present completely opposite statistics; while $\rho_{O}$ exhibits antibunching [$g_{O}^{(2)}(0)<1$], $\rho_{E}$ displays superbunching [$g_{E}^{(2)}(0)>2$]. To be more specific, $g_{E}^{(2)}(0)>2$ $\forall r\ge 0$ ($\lambda_r \ge 0$) and $g_{O}^{(2)}(0)<1$ for $0 \le \lambda_r < \sqrt{\sqrt{5}-2} \approx 0.49 \leftrightarrow 0 \le r < \tanh^{-1}\sqrt[4]{\sqrt{5}-2}\approx 0.86$. Figure ~\ref{fig:Second-order-correlation}(a) displays the change of $g^{(2)}(0)$ with $\lambda_r$ for $\rho_\text{th}$, $\rho_{E}$, $\rho_{O}$ and also for the single-mode squeezed state $\ket{\rm SS}=e^{\left(re^{-{\rm i}\theta}a^{2}-re^{{\rm i}\theta}a^{\dagger2}\right)/2}\ket{0}$, with $g_\text{SS}^{(2)}(0) = 2 + 1/\lambda_{r}$. We have included the latter for the sake of comparison, since it has the same average number of excitations of each single mode of the TMSS ($\moy{n}_\text{SS} = \moy{n}_\text{th}$) and presents superbunching. Remarkably, each mode of the even TMSS exhibits more superbunching than if it is in a single-mode squeezed state [$g_\text{E}^{(2)}(0)>g_\text{SS}^{(2)}(0)>2$] within the parameter range $0 \le \lambda_r < \sqrt{2}-1 \approx 0.41 \leftrightarrow 0 \le r < \tanh^{-1}\sqrt{\sqrt{2}-1}\approx 0.76$. 

Due to the aforementioned attributes, each mode of the even and odd TMSSs are quite suitable for quantum devices related to advanced imaging techniques and single-photon generation, respectively.
A single photon may be produced in each of the two modes when the odd TMSS is produced. The probability of projecting onto the odd TMSS is $\mathcal{P}_O = \lambda_r/(1+\lambda_r)$.
The proportion of the odd TMSS described by two single photons is $P_{1}^{O} = (1-\lambda_{r}^2)$.
Production of high purity states of two single photons requires $\lambda_r \rightarrow 0$, which comes at the expense of a low production probability $\mathcal{P}_O$, as shown in Fig.~\ref{fig:Second-order-correlation}(b). It is worth noting that the qubit may be used to herald projection onto the odd TMSS.

\section{\label{sec:4}Entanglement}
Considering a bipartite system in a pure state the degree of entanglement between the subsystems can be quantified through the linear entropy $E = \tfrac{d}{d-1}(1 - \gamma)$, where $\gamma=\Tr(\rho_{1}^2)$ is the purity of one of the subsystems described by the reduced density matrix $\rho_{1}$ and $d=\dim{\rho_{1}}$. Since our subsystems are bosonic modes, with infinite-dimensional Hilbert spaces, $0\le E \le 1$ such that $E=0$ for separable states while $E=1$ for maximally entangled continuous-variable states.

\begin{figure}[ht]
\includegraphics[trim = 18mm 5mm 16mm 7mm,clip,width=0.99\linewidth]{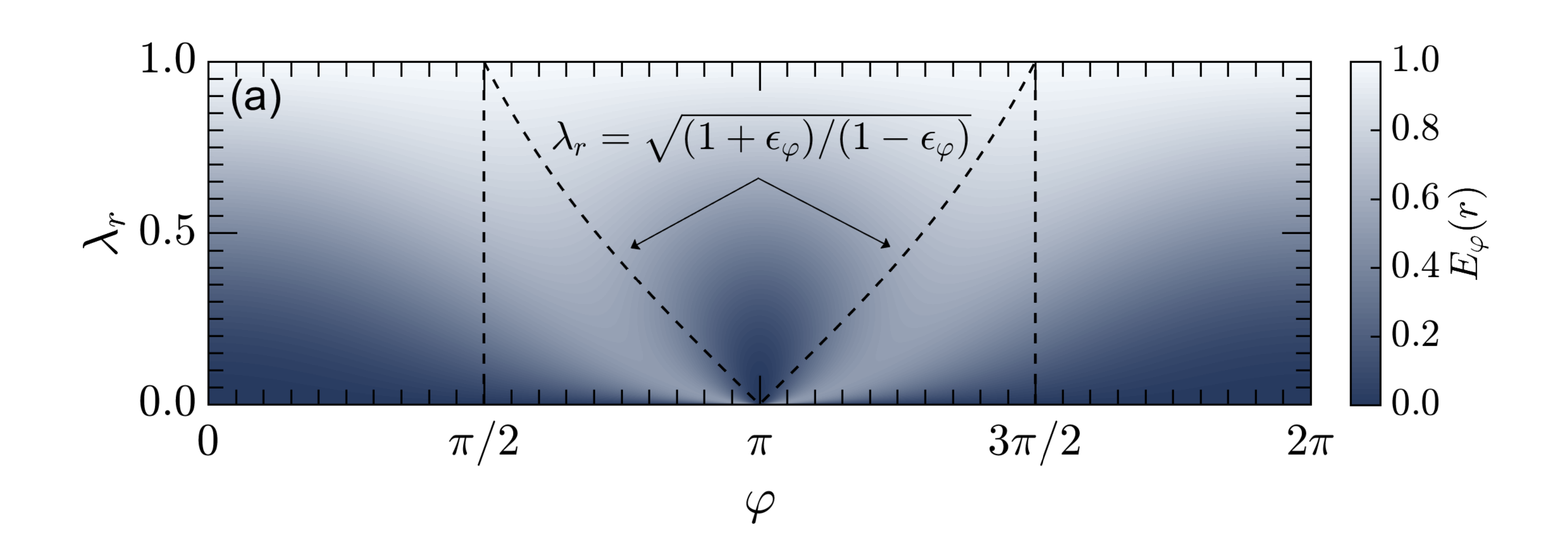} \\
\includegraphics[trim = 2mm 2.5mm 2mm 0mm,clip,width=0.495\linewidth]{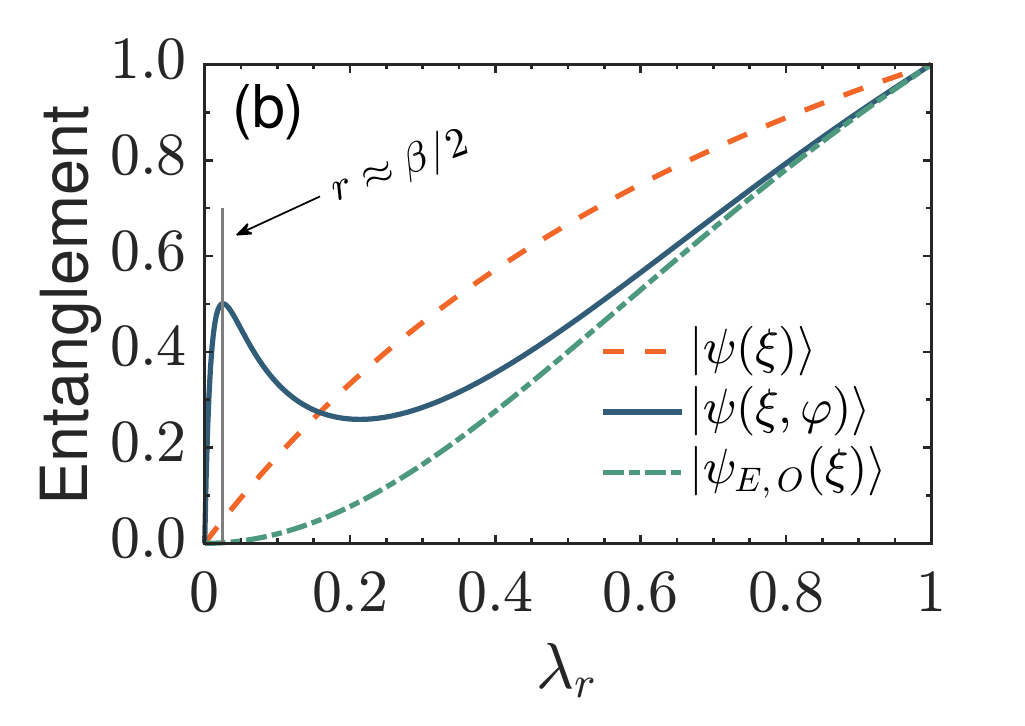}
\includegraphics[trim = 2mm 2.5mm 2mm 0mm,clip,width=0.495\linewidth]{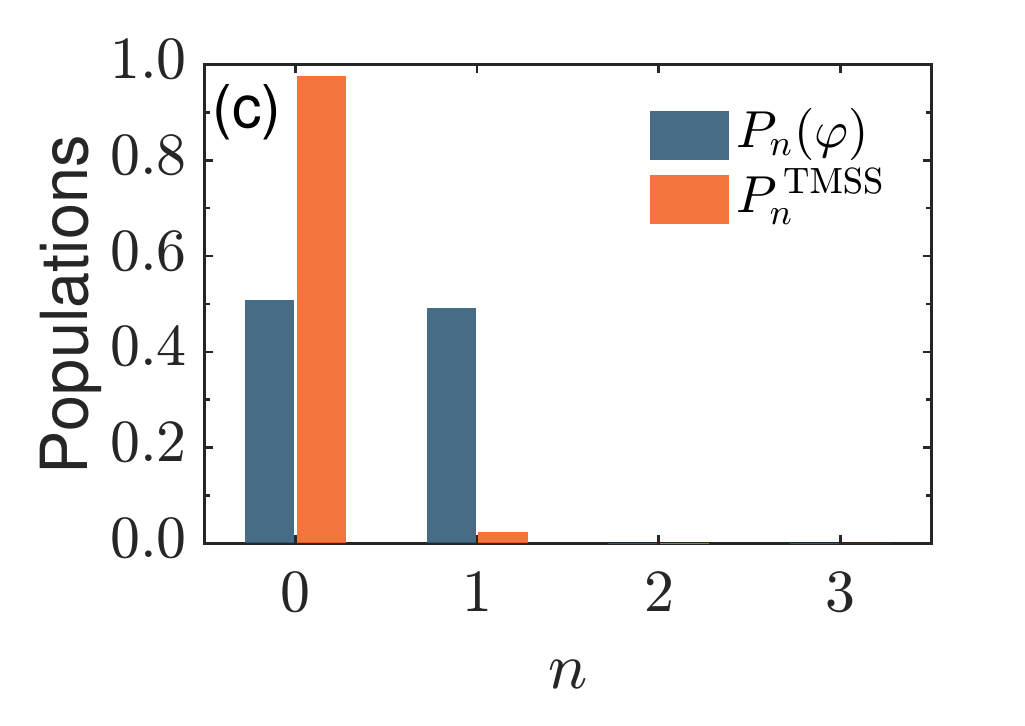}
\caption{\label{fig_entanglement} (a) Density plot showing the entanglement between the two modes [$E_{\varphi}(r)$], when they are in the cat-like state $\ket{\psi_{+}(r,\varphi)}$, as a function of the initial-state relative phase $\varphi$ and the two-mode squeezing parameter $r$ via $\lambda_r = \tanh^{2}(r)$. The dashed lines delimit the region for which $E_{\varphi}(r)$ surpasses the entanglement of the TMSS [$E_\text{TMSS}(r)$]. (b) Entanglement degree as a function of $\lambda_r$ considering $\varphi = \pi + \beta$, with $\beta = \pi/10$. One can notice here that the entanglement outperforming can occurs even for small values of $r$, reaching its maximum for $r \approx |\beta|/2$ ($\lambda_r \approx r^2$) when $|\beta| \ll 1$. Surprisingly, under this condition an extreme contrast between the entanglement in the cat-like state and in the TMSS takes place, while the latter is essentially a separable state given by the two-mode ground state ($\ket{\psi(\xi)} \approx (\ket{0,0} - (|\beta|/2) e^{\text{i}\theta}\ket{1,1})/\sqrt{1+|\beta|^{2}/4} \to E_\text{TMSS}(r) \approx 0.05$ for the considered paramaters), the former is approximately a maximally entangled qubit state ($\ket{\psi_{+}(\xi,\varphi)} \approx (\ket{0,0} - \text{i} e^{\text{i}\theta}\ket{1,1})/\sqrt{2} \to E_\varphi(r) \approx 0.5$). The even and odd TMSSs have the same entanglement degree, which is always smaller or equal than that for the TMSS. (d) Populations of the cat-like state ($P_{n}(\varphi)$) and the TMSS ($P_n^\text{TMSS}$) for $r = \beta/2$ and $\varphi = \pi +\beta$, with $\beta = \pi/10$.}
\end{figure}

For the TMSS $\ket{\psi(\xi)}$, the degree of entanglement is
\begin{equation}
E_\text{TMSS}(r)=1-\frac{1-\lambda_{r}}{1+\lambda_{r}}, 
\end{equation}
while for the general superposition $\ket{\psi_{+}(\xi,\varphi)}$, 
\begin{equation}
   E_{\varphi}(r) = 1-\left( \dfrac{1-\lambda_{r}^2}{1+\lambda_{r}^2}\right)\dfrac{(1+\epsilon_{\varphi})^2+\lambda_{r}^2(1-\epsilon_{\varphi})^2}{[(1+\epsilon_{\varphi})+\lambda_{r} (1-\epsilon_{\varphi})]^2}.
\end{equation}
$E_{\varphi}(r) = E_\text{TMSS}(r)$ for any $r$ when $\varphi=\{\tfrac{\pi}{2},\tfrac{3\pi}{2}\}$ $\to$ $\epsilon_{\varphi}=0$, which is not a surprise, since we have seen in Sec.~\ref{sec:3} that $\rho(r,\varphi)$ reduces to $\rho_\text{th}$ for these values of $\varphi$. The same degree of entanglement also occurs when $\lambda_r = \sqrt{(1+\epsilon_\varphi)/(1-\epsilon_\varphi)}$ or $r \to \infty$. Notably, $E_{\varphi}(r) > E_\text{TMSS}(r)$ for $\varphi \in (\tfrac{\pi}{2},\pi) \cup (\pi,\tfrac{3\pi}{2})$  provided that $0<\lambda_r < \sqrt{(1+\epsilon_\varphi)/(1-\epsilon_\varphi)}$. In this range, for the same value of the squeezing parameter
$r$, the entanglement degree in the TMSS cat-like states becomes higher than that in the TMSS. Curiously, it can reach high values even for $r \ll 1$, namely $E_{\varphi}(r) \approx 0.5$ for $\varphi = \pi + \beta$ and $r\approx |\beta|/2$ considering $|\beta| \ll 1$, for which we have $\ket{\psi_{+}(\xi,\varphi)} \approx (\ket{0,0} - \text{sgn}(\beta) e^{\text{i}(\theta + \tfrac{\pi}{2})}\ket{1,1})/\sqrt{2}$, i.e., a maximally entangled qubit state. By contrast, $E_\text{TMSS}(r) \approx \beta^{2}/2 \ll 1$ under the same conditions, for which we have $\ket{\psi(\xi)} \approx (\ket{0,0} - (|\beta|/2) e^{\text{i}\theta}\ket{1,1})/\sqrt{1+|\beta|^{2}/4} \approx \ket{0,0}$, i.e., a separable state. This means that it is possible to generate much more entanglement between the modes with a lower squeezing parameter by exploiting the cat-like states instead of the TMSS, indicating an advantage from the point of view of quantum information science. Figure \ref{fig_entanglement} illustrates the above results.
It is worth noting that the even ($\varphi=0)$ and odd ($\varphi=\pi)$ TMSSs have the same entanglement degree, $E_{E}(r) = E_{O}(r) =  1- (1-\lambda_{r}^2)/(1+\lambda_{r}^2)$, which is always smaller or equal than that for the TMSS. Despite that, we show in the next section that their RSMSs can be employed for quantum metrological purposes.

\section{\label{sec:5}Quantum Metrology}

Quantum metrology takes advantage of the properties of quantum mechanics
to better estimate parameters involved in dynamical processes, using quantum
states as probes \cite{helstrom1969}. The process of estimating a parameter $y$ follows a specific sequence of steps, known as
the protocol of estimation \cite{Escher2011,*escherdavidovich2011}: (i)~the probe
state is prepared in an initial and determined configuration, represented
by the density matrix $\rho$; (ii)~the initial state evolves through
a dynamical process, which is represented by a unitary evolution
operator $U(y)$, such that the final configuration of the system is dependent
on the parameter $y$; (iii)~the final state $\rho(y)$ is measured, giving results $y_\text{est}(\kappa)$, with associated
probabilities $P_{\kappa}(y)$; (iv)~these results are used to estimate the parameter $y$. It is important
to note that different results $\kappa$ come from separated processes
of measurement. The average value of the parameter is then calculated
with the individual estimation $y_\text{est}(\kappa)$ as $\moy{y_\text{est}}=\sum_{\kappa}y_\text{est}(\kappa)P_{\kappa}(y)$,
with $\sum_{\kappa}P_{\kappa}(y)=1$. The deviation on the parameter is defined as $(\Delta y)^{2}\text{\ensuremath{\equiv}}\moy{(y_\text{est}-\moy{y_\text{est}})^{2}}$ and its lower bound is proportional to the inverse of the Fisher information $ \moy{\left(\frac{\mathrm{d\ln(P_{\kappa}(y))}}{\mathrm{d}y}\right)^{2}}$ \cite{fisher1922,fisher1925theory} with respect to parameter $y$,
and represents the maximum quantity of information that can be obtained
with respect to $y$ through the probability set $\left\{ P_{\kappa}(y)\right\} $, as derived from the Cramér-Rao inequality \cite{cramerstatistics1946,rao1973}.

Given a generator $U(y)$ for the transformation on the density matrix
which leads to the final state $\rho(y)$, the quantum Fisher information
for $y$ with probe state $\rho$ can be written as $F(y)=4[\Delta U(y)]^{2}$
if $\rho$ is a pure state, with $[\Delta U(y)]^{2}=\moy{U^{2}(y)}-\moy{U(y)}^{2}$,
and $F(y)=2\sum_{i,j}\frac{(\lambda_{i}-\lambda_{j})^{2}}{\lambda_{i}+\lambda_{j}}\abs{\bra{i}U(y)\ket{j}}^{2}$
if $\rho$ is mixed, where $\lambda_{j}$ and $\ket{j}$ stand for
the eigenvalues and eigenvectors of $\rho$, respectively.

Single-mode squeezed states are widely used in quantum metrology due to the reduction of fluctuations in one of the quadratures of the bosonic mode \cite{caves1981,pezze2008mach,sahotaquesada2015}.
The fact that the Wigner functions of each mode of the even and odd TMSSs are invariant under rotations and sharply concentrated around the origin of the phase space (Fig.~\ref{Pn_Wigner}) motivates the study of applications of
those states in quantum metrology.

\begin{figure}[b]
\includegraphics[trim = 2mm 0mm 7mm 12mm,clip,width=0.6\linewidth]{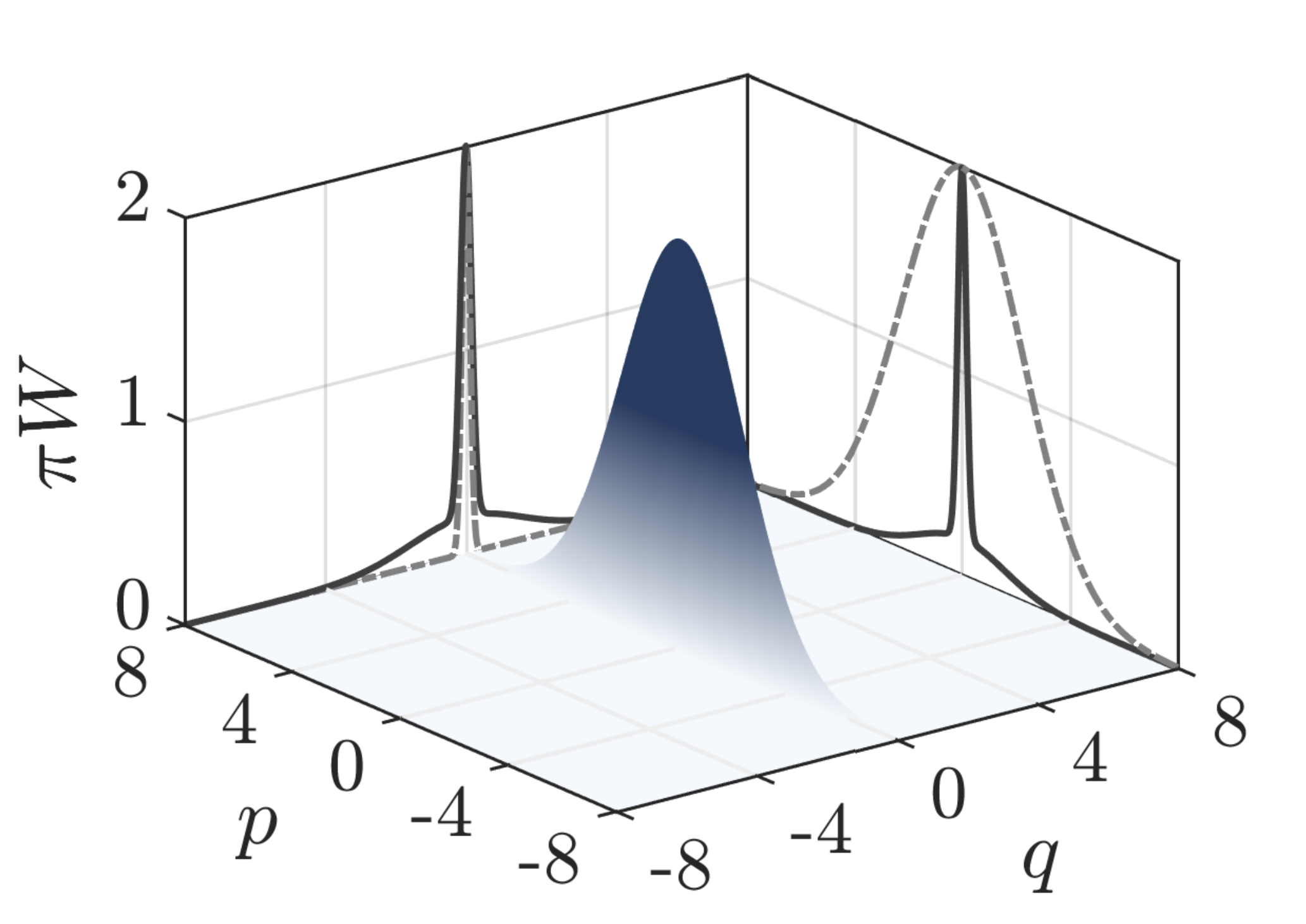}
\caption{\label{fig:wigner-comparison}The Wigner function (and its projections)
of the single-mode squeezed state is compared with the projections of the RSMS of the Even TMSS,
with $r=1.5$. The projections of the RSMS of the Even TMSS display tails because they are described by the summation of two two-dimensional Gaussian functions (see Eq.~\eqref{eq: W_even_odd}).}
\end{figure}

\begin{figure}[t]
\includegraphics[trim = 2mm 2.5mm 2mm 0mm,clip,width=0.495\linewidth]{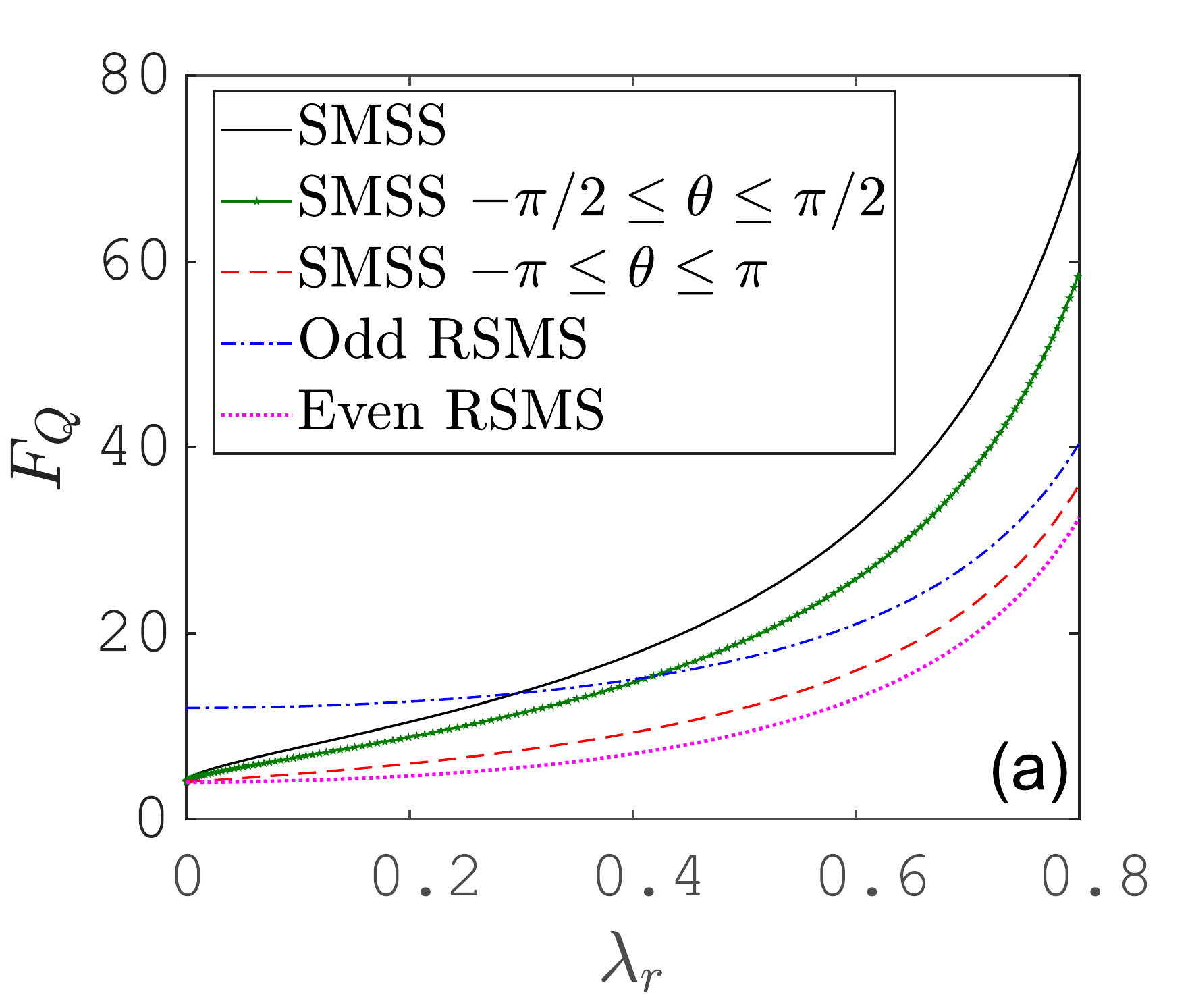}
\includegraphics[trim = 2mm 2.5mm 2mm 0mm,clip,width=0.495\linewidth]{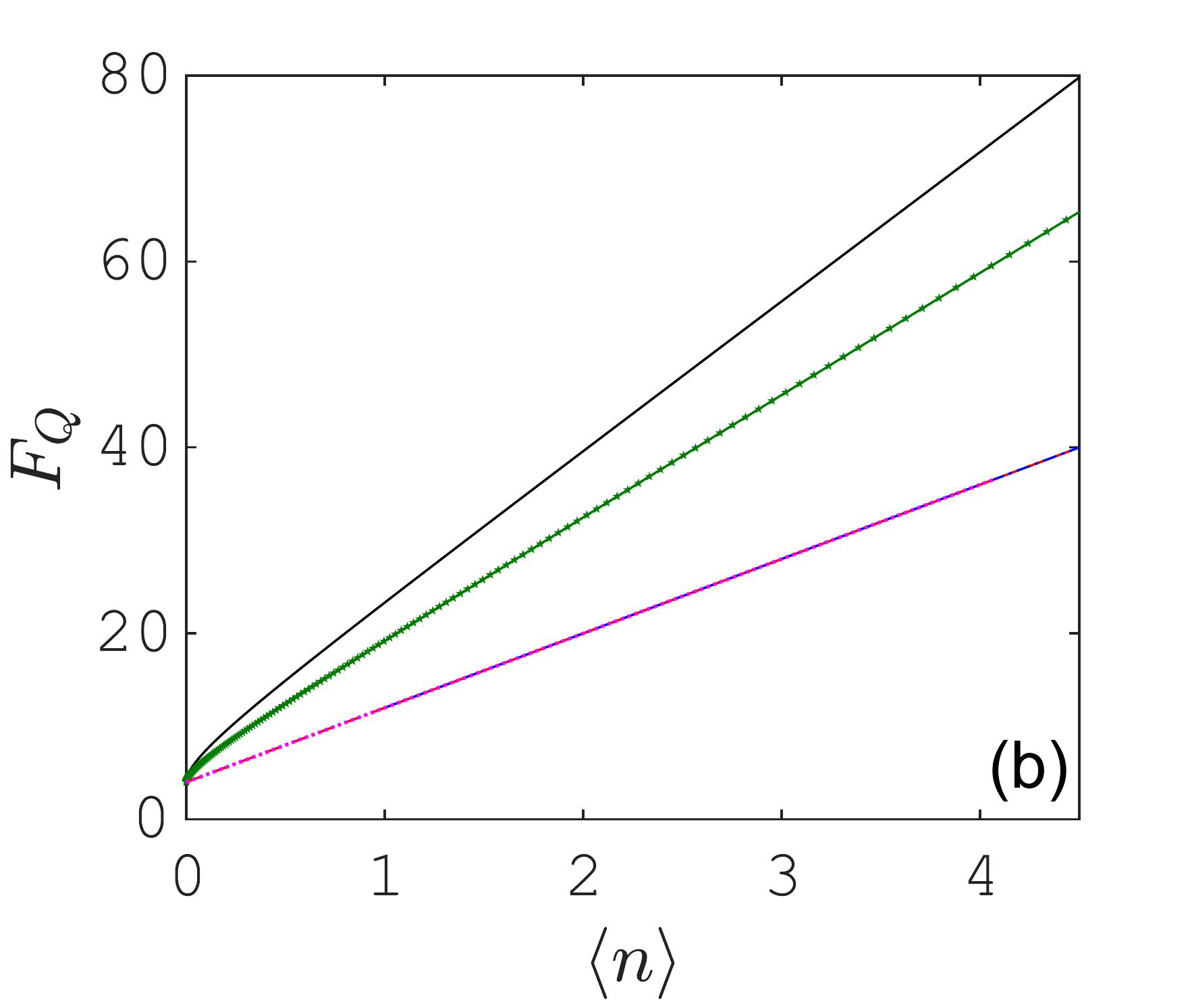}
\caption{\label{fig:qfi}Quantum Fisher Information $F_Q$ as a function of $\lambda_r$ (a) and the average number of excitations in the bosonic field $\expval{n}$ (b) for each state. If the direction of squeezing in phase space matches the displacement direction, single-mode squeezed states (SMSS) are well-suited for measuring the displacement.
This is shown by the quantum Fisher information $F_Q$ of the black curve.
If the angle $\theta$ between the squeezing direction and the displacement direction is large, it can be more efficient to measure the displacement using RSMS of the even or odd TMSS.
This is seen by comparing the red curve (the red and green curves show the average $F_Q$ values obtained with SMSSs when $\theta$ is spread over a range) with the blue and pink curves.
$\lambda_r = \tanh^2{r}$ describes the amount of squeezing.
Note also for small values ($\lambda_r<0.3$), the odd RSMS presents a larger $F_Q$ than the SMSS. In particular, for $\lambda \rightarrow 0$, the odd RSMS approximates to the single-photon state $\ket{1}$, and the SMSS to the vacuum $\ket{0}$. From (b), it is seen that $F_Q$ behaves linearly for all states, and, in particular, for small values of $\lambda$, $F_Q(\expval{n}) \approx 4(2\expval{n} + 1)$, with $\expval{n}$ being the average number of excitations for each corresponding state. 
}
\end{figure}

Figure~\ref{fig:wigner-comparison}
shows a comparison between the Wigner functions of a single-mode squeezed state, with reduced noise in one of the quadratures, and the
RSMS of the even TMSS, for the same value of the squeezing parameter $r$. The single-mode squeezed
state has a more concentrated probability distribution profile
than the RSMS. However, we would like to point out that the
symmetry of the Wigner function of the even and odd RSMSs implies a robust scheme for applications in quantum metrology. As the even and odd RSMSs exhibit squeezing characteristics in
all directions around the origin of the phase space, they represent a useful class of
probe states for detecting small displacements in any
direction. Furthermore, the Wigner function still presents thermal contributions along broad regions of the phase space, so, despite the concentrated peak around the origin, the Heisenberg uncertainty principle is not violated. For estimation of displacement, we employ
the general quadrature operator $X(\phi)=a\mathrm{e}^{-\I\phi}+a^{\dagger}\mathrm{e}^{\I\phi}$,
which corresponds to position and momentum operators for $\phi=0,\pi/2$,
respectively. In Fig.~\ref{fig:qfi} are represented the quantum
Fisher information for the RSMSs of the even and odd TMSSs and for the single-mode squeezed
state, for estimation of position amplitudes of the mode.

Although the single-mode squeezed state shows better results for the quantum Fisher information (for a specific quadrature), by employing a measurement scheme of the Wigner function \cite{Lutterbach1997}, the even and odd RSMSs could be employed to detect small coherent forces.

For our state, differently from the single-mode squeezed state, one does not need to worry about the phase of the displacement due to the symmetry of its Wigner function around the origin, as we see in Fig. \ref{fig:wigner-comparison}. In the next section we describe in detail how our sensor of small forces can be implemented in trapped-ion systems.

\section{\label{sec:6}Ion implementation}

TMSSs can be generated by a combination of Jaynes-Cummings and anti-Jaynes-Cummings interactions, as we see in Eq.~\eqref{eq:heff4}. In this section we describe how these interactions can be realized to produce an entangled state of two motional modes of a single trapped ion, which could be used for enhanced force sensing. In Ref. \cite{Zeng2002} similar effective Hamiltonians are proposed, but the proposal involves two trapped ions and other types of two-mode squeezed states are considered.

Electronic states of a single trapped ion can be coupled to the ion's motion using a laser field. When the laser field's wavevector projects onto two motional modes ($x$ and $y$), the Hamiltonian describing the coupling (within the interaction picture and after taking the rotating-wave approximation) is
\begin{equation}\label{eq: H_full}
H_F = \frac{\Omega}{2}\mathrm{e}^{-\I \Delta t}  \mathrm{e}^{\I \eta_x (a \mathrm{e}^{-\I \omega_x t} + a^{\dag} \mathrm{e}^{\I \omega_x t})} \mathrm{e}^{\I \eta_y (b \mathrm{e}^{-\I \omega_y t} + b^{\dag} \mathrm{e}^{\I \omega_y t})} \sigma^+ + \mathrm{H.c.}
\end{equation}
where $\Omega$ is the coupling strength, $\Delta$ is the detuning of the laser field from the atomic resonance, $a^{\dag}, b^{\dag}$ and $a,b$ raise and lower the states of the $x$ and $y$ modes, and $\omega_i$ are the motion mode frequencies. $\sigma^+$ and $\sigma^-$ act on the ion's internal state and the Lamb-Dicke parameters are defined by
\begin{equation}
\eta_i = k_i \sqrt{\frac{\hbar}{2 m \omega_i}},
\end{equation}
where $k_i$ is the projection of the laser field's wavevector in the $i$-direction and $m$ is the ion mass.

When a two-colour coupling field satisfying $\Delta = \pm (\omega_x + \omega_y)$ is used, and provided the system is within the Lamb-Dicke regime $\eta_x^2(2n_x+1), \eta_y^2 (2n_y+1) \ll 1$ ($n_i$ is the number of phonons in the $i$ mode) then the coupling Hamiltonian becomes
\begin{equation}\label{eq: H_eff}
\begin{split}
H_{\textrm{eff}} &=-\frac{1}{2} \eta_x \eta_y  \Omega \sigma^+ \left( a b + a^{\dag} b^{\dag} \right) + \mathrm{h.c.} \\
&=-\frac{1}{2} \eta_x \eta_y  \Omega \sigma_x \left( a b + a^{\dag} b^{\dag} \right)
\end{split}
\end{equation}
after another application of the rotating wave approximation.
Identifying $\chi = \frac{1}{2} \eta_x \eta_y \Omega$, this Hamiltonian is equivalent to Eq.~\eqref{eq:heff4} for real-valued $\chi$.

\begin{figure}[b]
\includegraphics[scale=0.35]{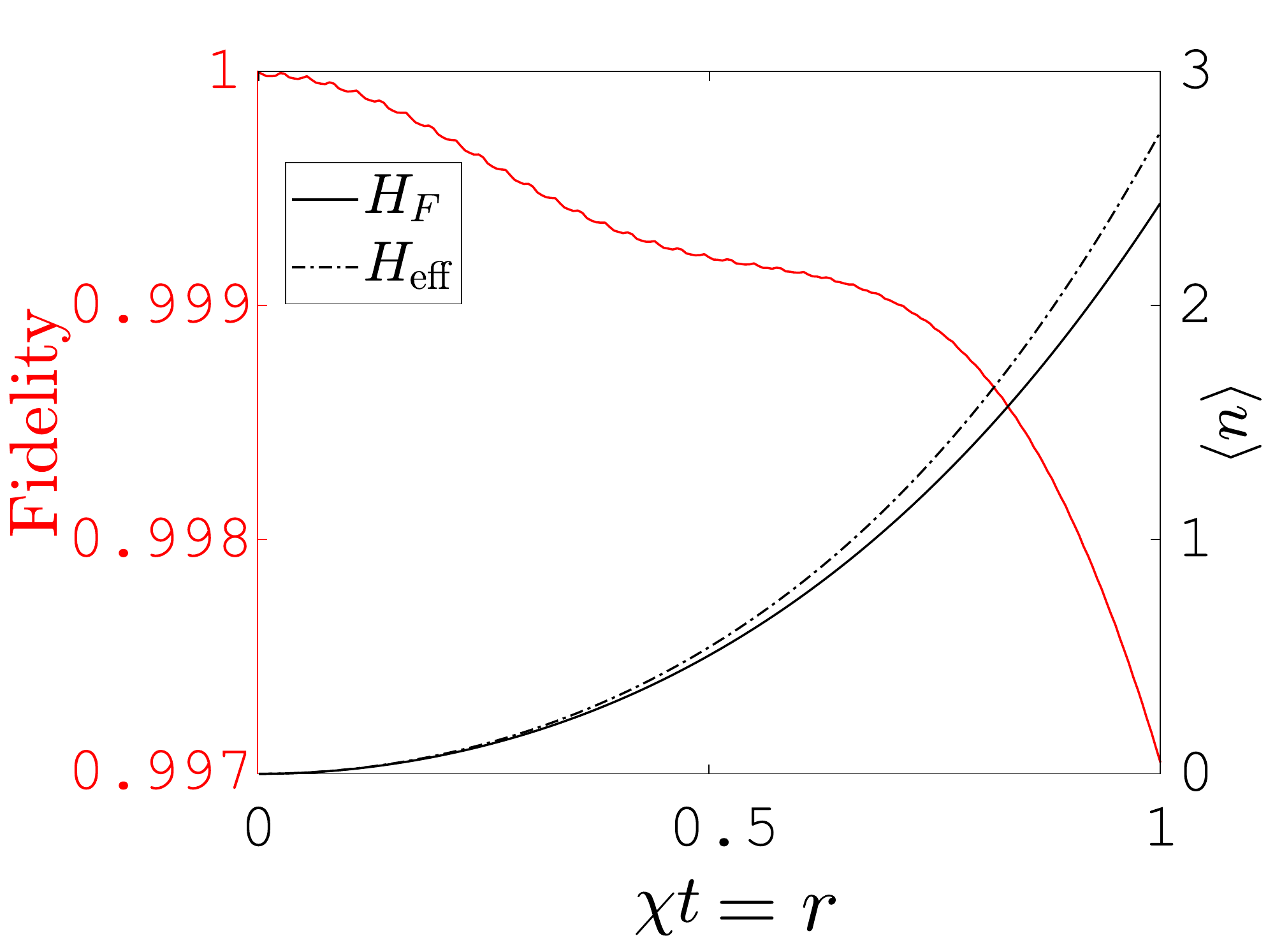}
\caption{\label{fig: fidelity}The squeezing dynamics in a trapped ion system is captured well by $H_\mathrm{eff}$: The fidelity (in red) shows the overlap of the state evolved according to the effective Hamiltonian $H_\mathrm{eff}$ with the state evolved according to the full Hamiltonian $H_F$. The evolution of the average number of excitations in the bosonic fields is also shown.}
\end{figure}

The coupling laser field drives second-order sideband transitions which are relatively weak.
This implementation requires the coupling dynamics to be faster than the decoherence time $\eta_x \eta_y \Omega \gg \gamma$, also off-resonant excitation of stronger transitions must be avoided $\Omega \ll \omega_x + \omega_y$.

If the system is initialized in $\ket{\Psi(t=0)} = \ket{g} \ket{0,0}$, then after evolution under $H_F$ given by Eq.~\eqref{eq: H_eff}, followed by projection of the electronic state onto $\ket{g}$ ($\ket{e}$) the motional modes of the trapped ion can be prepared in the even (odd) TMSS in principle.
Projective measurement of a trapped ion's electronic state is commonly accomplished with near-unity fidelity by detecting laser-induced fluorescence  detection \cite{Leibfried2003}.
We note that if the ion is in state $\ket{g}$, the scattering of fluorescence photons destroys the ion's motional state.
If the ion is in projected onto the non-fluorescing state ($\ket{e}$) the motional state (the odd TMSS) will be unperturbed.

In Fig. \ref{fig: fidelity}, we plot the fidelity of the states which evolve under the effective Hamiltonian $H_{\textrm{eff}}$ (Eq.~\ref{eq: H_eff}) as compared with the states evolved under the full Hamiltonain  $H_F$ (Eq.~\ref{eq: H_full}).
The effective Hamiltonian describes the squeezing dynamics well up to $\chi t = r = 1$.
The parameters considered were $\omega_x=1.0$, $\omega_y=1.2$, $\Omega = \omega_x/20$, $\eta_x=\eta_y=0.1$.
We also show the evolution of the average number of phonons $\expval{n}=a^{\dagger}a+b^{\dagger}b$.

Now we describe how these states can be employed to detect small forces by measuring the Wigner function of one of the modes based on the protocol given in Ref. \cite{Lutterbach1997}. By tracing out one of the modes, one can see that the Wigner function of the reduced state is completely symmetric around the origin (see Fig. \ref{fig:wigner-comparison}). Thus, one does not need to worry about the phase of the coherent displacement applied onto the ion trap.

A weak coherent force applied on either the ion or the ion trap causes a displacement $D(\alpha) = e^{\alpha a^{\dagger} -\alpha^{*} a}$ of the mode state, resulting in a state $\rho(\alpha)= D(\alpha)\rho_v D^{-1}(\alpha)$, where $\rho_v$ is the reduced density operator of the motional state. The phase and amplitude of the complex parameter $\alpha$ indicate the direction and intensity of the displacement operation in the phase space. At this point a laser pulse resonant to the $\ket{g} \leftrightarrow \ket{e}$ transition (the carrier transition) is applied, whose Hamiltonian is given by (keeping terms up to $\eta_x^2$) $H_c=\Omega/2 [1-\eta_x^2( a^{\dagger} a +1/2)]\sigma_x$ . Adjusting the interaction time $\tau$ such that $\Omega\eta_x^2\tau/2= \pi/2$, the evolution operator will be given by $U=e^{-i H_c \tau} = e^{-i (\Phi - \pi a^{\dagger}a/2)} $, with $\Phi = \Omega \tau/2 - \pi/4$. As shown in \cite{Lutterbach1997}, the evolved state of the system, $ U\rho(\alpha) \ket{e} \bra{e}U^{-1}$, will be 
\begin{align}
\begin{split}
    \left[\ket{e}\cos(\Phi-\pi a^{\dagger}a/2) - i\ket{g}\sin(\Phi-\pi a^{\dagger}a/2)\right] \rho(\alpha)\\
    \times  \left[\bra{e}\cos(\Phi-\pi a^{\dagger}a/2) +i\bra{g}\sin(\Phi-\pi a^{\dagger}a/2)\right]
\end{split}
\end{align}
and then the population inversion $P_{eg} = \langle \sigma_z \rangle$ of the ion will give, apart from a constant factor, the value of the Wigner function at the position $\alpha = (q-ip)/2$ in phase space, i.e., $P_{eg} \propto W(\alpha)$ \cite{Lutterbach1997}. For the even (odd) TMSS, when there is no force acting on the ion this results in $\alpha=0$ and, consequently, the maximal (minimal) value of the atomic population inversion. However, for small values of $|\alpha|$ -- which are larger than the width of the central peak of the Wigner function -- the population inversion would result in nearly zero, thus allowing us to indiscriminately detect the action of small coherent forces.

\section{\label{sec:7}Conclusions}

In this work, we present the statistical properties of superpositions of TMSSs (two-mode squeezed states) with relative phase factors, giving special attention to two cases of relative phase, corresponding to the even and odd TMSSs. The RSMSs (reduced single-mode states) of the even ($\rho_E$) and odd ($\rho_O$) TMSSs, obtained by tracing out one of the bosonic modes, present populations in the Fock basis which resemble thermal distribution, thus illustrating the ``pseudo-thermal" behavior of those states. Furthermore, we investigate the second-order correlation function of $\rho_E$ and $\rho_O$, and show that, for small squeezing parameters, $\rho_E$ presents superbunching behavior, while $\rho_O$ presents antibunching, thus making it a potential source of single photons. We also study entanglement between the two bosonic modes corresponding to the superposition of TMSSs, and, for small values of the squeezing parameter $r$ and specific relative phase angle $\varphi$, the superpositions show a larger degree of entanglement than TMSSs, generating a maximally mixed state in each of the RSMSs when tracing out one of the mode variables. We also study the pseudo-probability distributions related to $\rho_E$ and $\rho_O$ in phase space, given by the Wigner function $W(q,p)$. We point out that the profiles of both RSMSs narrow around the phase space origin as the squeezing parameter is increased. For $\rho_E$ and $\rho_O$, the symmetry of $W(q,p)$ makes the states sensible to weak forces in any direction of the phase space, in contrast with the single-mode squeezed state. Finally, we describe how the states discussed above can be generated and how they can be employed to measure weak forces in the trapped-ion domain, by deriving effective two-mode Jaynes-Cummings-like interactions, thus motivating applications in quantum information processing, quantum metrology and quantum sensing of small coherent displacements.


\begin{acknowledgments}
This work was supported by the Coordenação de Aperfeiçoamento de Pessoal de  Nível Superior (CAPES)
- Finance Code 001, and through the CAPES/STINT project, grant No.
88881.304807/2018-01. C.J.V.-B. is also grateful for the
support by the São Paulo Research Foundation (FAPESP)
Grant No. 2019/11999-5, and the
National Council for Scientific and Technological Development (CNPq) Grant No. 307077/2018-7. This work is also part of the Brazilian National Institute of Science and Technology for Quantum Information
(INCT-IQ/CNPq) Grant No. 465469/2014-0. G.H.\ gratefully acknowledges the hospitality of Universidade Federal de São Carlos (UFSCar) and Universidade de São Paulo (USP).
\end{acknowledgments}

\bibliography{refs1}

\end{document}